\documentclass[%
 aip,
 jmp,%
 amsmath,amssymb,
reprint,%
]{revtex4-1}

\usepackage{graphicx}
\usepackage{dcolumn}
\usepackage{bm}

\usepackage{color} 

\begin{document}

\preprint{AIP/123-QED}

\title[An ab initio approach to free-energy reconstruction]
{An ab initio approach to free-energy reconstruction using logarithmic mean force dynamics}


\author{Makoto Nakamura}
\email{nakamura@cphys.s.kanazawa-u.ac.jp}
\affiliation{Graduate School of Natural Science and Technology, Kanazawa University, 
Kanazawa 920-1192, Japan}
\author{Masao Obata}%
\affiliation{Graduate School of Natural Science and Technology, Kanazawa University, 
Kanazawa 920-1192, Japan}

\author{Tetsuya Morishita}
\affiliation{Nanosystem Research Institute,
National Institute of Advanced Industrial Science and Technology (AIST), 
Tsukuba 305-8568, Japan}%
%
\author{Tatsuki Oda}
\email{oda@cphys.s.kanazawa-u.ac.jp}
\affiliation{Graduate School of Natural Science and Technology, Kanazawa University, 
Kanazawa 920-1192, Japan}
\affiliation{Institute of Science and Engineering, Kanazawa University, 
Kanazawa 920-1192, Japan}
%

\date{\today}

\begin{abstract}
We present an {\it ab initio} approach for evaluating a free energy 
profile along a reaction coordinate
by combining logarithmic mean force dynamics (LogMFD) and
first-principles molecular dynamics.
The mean force, which is the derivative of the free energy with respect to
the reaction coordinate, is estimated using 
density functional theory (DFT) in the present approach,
which is expected to provide an accurate free energy profile
along the reaction coordinate.
We apply this new method, first-principles LogMFD (FP-LogMFD),
to a glycine dipeptide molecule and reconstruct one- and two-dimensional
free energy profiles in the framework of DFT.
The resultant free energy profile is compared with that obtained by the thermodynamic
integration method and by the previous LogMFD calculation using an empirical force-field,
showing that FP-LogMFD is a promising method to calculate free energy 
without empirical force-fields.
\end{abstract}


\keywords{mean force dynamics, free energy, logarithmic energy potential, 
               glycine dipeptide, 2-(Acetylamino)-N-methylacetamide}

\maketitle

\section{\label{sec:introduction}{INTRODUCTION}}

Free energy is a significant physical property for estimating thermodynamic stability.
It is desirable to estimate free energy as accurately as possible. 
Such free energy estimation is becoming important in a variety of research fields; 
in particular, biological molecules including proteins or interfaces of nano-scale materials 
have been raised as a target for free energy calculations\textcolor{black}{\cite{Yamamori2013,Sinko2012,
Konig2014}}. 
It is thus desirable to develop methods that improve the accuracy and efficiency
in free energy calculations using molecular simulation.
The free energy in molecular systems has often been evaluated for a given constraint 
(reaction pass).\cite{Ciccotti2004}
Such a constraint is usually specified by using a set of reaction 
coordinates,\cite{Sprik1998} for example, distances 
between molecules, bond angles, and dihedral angles, etc.

In order to get free energy landscapes, various techniques 
[thermodynamic integration (TI)\cite{Kirkwood1935}, 
 free energy perturbation,\cite{Zwanzig1954} 
 umbrella sampling,\cite{Torrie1977} and so on] have been developed so far.  
Although TI is derived from the statistical mechanics faithfully, 
some difficulties have been pointed out;
poor sampling which could come from a breakdown of the ergodicity and
numerical integration as postprocessing.

To overcome these difficulties, free energy calculation methods
based on mean force dynamics (MFD) have been proposed.\cite{Rosso2002,Laio2002}
In MFD, a set of $L$ reaction coordinates (collective variables),
${\bf X} \equiv \{X_{1}, X_{2}, \cdots, X_{L} \}$, is regarded as a set of fictitious
dynamical variables, and their trajectories are designed to be generated
by hypothetical dynamical equations.
Morishita {\it et al.}\cite{Morishita2012,Morishita2013}
have recently introduced a logarithmic form of the free energy along ${\bf X}$ [$F({\bf X})$]
to enable us to easily sample rare events in MFD calculations. 
This method is called logarithmic mean force dynamics (LogMFD),
in which the free energy can be estimated on-the-fly.

The evaluation of mean force (MF), i.e., slope of  $F({\bf X})$ with respect to ${\bf X}$,
in such a method based on MFD can be 
improved 
by incorporating first-principles (FP) molecular dynamics 
(MD),\cite{Car1985} replacing the classical MD using empirical force fields.
FPMD allows us to include effects of the electronic state explicitly; 
for example, bond-formation or bond-breaking, 
which may considerably influence the free energy profiles in molecular systems.

In this paper, 
we have developed 
first-principles MFD in the framework of LogMFD, namely,
first-principles LogMFD (FP-LogMFD).
We reconstruct the free energy landscape 
for a molecular system of glycine dipeptide using FP-LogMFD. 
This demonstration indicates that FPMD can be 
incorporated into LogMFD of
multi-dimensional ${\bf X}$-systems
and that the scheme developed here is found to be promising 
for the free energy reconstruction using {\it ab initio} techniques.  
The successful combination of LogMFD and FPMD is indebted to 
the efficiency for sampling rare events in LogMFD. The logarithmic form 
introduced in LogMFD suppresses the effective energy barriers for the 
dynamical variables ${\bf X}$. 
This makes it possible to sample configurations with higher energy, 
as frequently as those with much lower energy. 
This feature also makes it possible 
to improve the accuracy of the MF by increasing the number of statistical samples (FPMD steps).

In the next section, we review the LogMFD method briefly and 
demonstrate how to incorporate FPMD into LogMFD. 
In Sec. III, we present the free energy profile with respect 
to the dihedral angles in glycine dipeptide molecule. 
In Sec. IV, we will discuss the entropic contribution and numerical accuracy
in the present results
by comparing it
with the \textcolor{black}{classical MD}
result previously obtained using
an empirical force field. Finally, Sec. V summarizes this paper.

\section{\label{sec:theory}Methods}

\subsection{\label{subsec:theory}Equations of mean force dynamics}

We present a brief review for LogMFD. 
This review would be a good introduction to our new scheme which employs a non-empirical 
approach. 
We consider a system of $N$ atoms with a given temperature $T_{\rm ext}$,
and aim to reconstruct the free energy profile $F({\bf X})$ with respect to ${\bf X}$.
Each reaction coordinate $X_{p}(\{ {\bf R}_{I} \})$ is generally a function of 
the atomic coordinates $\{ {\bf R}_{I}\}$, 
where $p$ and $I$ specify the $p$'th reaction coordinate and 
the $I$'th atom, respectively. 
In MFD, however, ${\bf X}$ are regarded as dynamical variables, being independent of $\{ {\bf R}_{I}\}$.
We now consider the following postulated Hamiltonian for ${\bf X}$;

\begin{eqnarray}
H_{\rm MFD} & = & 
\sum_{p}^{L}\frac{1}{2}M_{p} \dot{X}_{p}^{2} + F({\bf X}), 
\label{HMFD}
\end{eqnarray}
where the first and second terms on the right-hand-side are the kinetic 
and potential energies, respectively, for $X_{p}$ ($\dot{X}_{p}$ means the 
velocity $d X_{p}/dt$) and $M_{p}$ is the fictitious mass for $X_{p}$.
The equation of motion for ${X_{p}}$ is thus obtained as,

\begin{eqnarray}
M_{p} \ddot{X}_{p} &=& -
                         \frac{\partial F({\bf X})}{\partial X_{p}}, 
\label{eqmf}
\end{eqnarray}
where $-{\partial F({\bf X})/\partial X_{p}}$ is the MF.
The solution for this equation of motion fulfills the conservation law, i.e., 
$H_{\rm MFD}$ can be seen as a constant of motion, as long as the MF
 is accurately evaluated.  

Several methods based on MFD have been proposed thus far, which provide
us free energy profiles with respect to reaction coordinates and allow us 
to discuss many kinds of physics involving the reaction coordinates.
Metadynamics\cite{Laio2002}  has been introduced utilizing the concept of MFD, 
and has been applied to a variety of systems including biosystems 
to sample rare events and to reconstruct free energy profiles.
Morishita {\it et al.}\cite{Morishita2012,Morishita2013} 
have proposed LogMFD in which $F({\bf X})$ in Eq. (\ref{HMFD})
is replaced with a logarithmic form of $F({\bf X})$, and have demonstrated 
several improvements in the free energy calculation.

In LogMFD,
the following Hamiltonian is introduced instead of Eq. (\ref{HMFD});
\begin{eqnarray}
H_{\rm LogMFD} & = & \sum_{p}^{L} \frac{1}{2} M_{p}\dot{X}_{p}^{2} + 
                  \gamma{\mathrm {log}} \{ \alpha F({\bf X})+ 1 \},
\label{HMFD1}
\end{eqnarray}
where $\gamma$ and $\alpha$ are positive constant parameters, 
which are chosen to effectively reduce the energy barriers experienced 
by ${\bf X}$. The resultant equation of motion for $X_{p}$ is given as,
\begin{eqnarray}
M_{p} \ddot{X}_{p} &=& - \left(\frac{\alpha \gamma}{\alpha F({\bf X})+1}\right)
                         \frac{\partial F({\bf X})}{\partial X_{p}}.
\label{eqmfdeta0}
\end{eqnarray}
In practice, ${\bf X}$ can be thermostatted in LogMFD calculations, 
and the equation of motion is slightly modified as follows;
\begin{eqnarray}
M_{p} \ddot{X}_{p} &=& - \left(\frac{\alpha \gamma}{\alpha F({\bf X})+1}\right)
                         \frac{\partial F({\bf X})}{\partial X_{p}}
                          - M_{p} \dot{X}_{p} \dot{\eta}, 
\label{eqmfd} \\
Q_{\eta} \ddot{\eta} &=&  \sum_{p} M_{p} \dot{X}_{p}^{2} -
                                  L k_{\rm B} T_{X},
\label{eqmfdeta}
\end{eqnarray}
where $\eta$ is the thermostat variable which controls the temperature of 
${\bf X}\ $\textcolor{black}{
($T_{X}$)},
 $Q_{\eta}$ is the mass for $\eta$,  
and $k_{\rm B}$ is Boltzmann's constant. 
With a single Nos\'e-Hoover thermostat \cite{Nose1984,Hoover1985} as in Eqs. (\ref{eqmfd}) and (\ref{eqmfdeta}), 
the following pseudo Hamiltonian is a constant of motion instead of $H_{\rm LogMFD}$;\cite{Morishita2012,Morishita2013}
\begin{eqnarray}
\hat{H}_{\rm LogMFD} & = & \sum_{p}^{L} \frac{1}{2} M_{p}\dot{X}_{p}^{2} + 
                  \gamma{\mathrm {log}} \{ \alpha F({\bf X})+ 1 \}
\nonumber \\
            &   & +\frac{1}{2}Q_{\eta} \dot{\eta}^{2}+ L k_{\rm B}T_{X} \eta.
\label{HMFD2}
\end{eqnarray}
Note that $T_{X}$ is not necessarily the same as
the temperature for atoms, $T_{\rm ext}$. 
The heights of the energy barriers on 
$\gamma \log \{ \alpha F({\bf X})+1 \}$ 
are much lower than those on $F({\bf X})$.
This reduction of the barrier height enables the coordinate $X_{p}$ 
to easily cross  the barriers at a moderate temperature of $T_{X}$, 
allowing us to evaluate the free energy associated with rare events.

$\partial F({\bf X})/\partial X_p$ is obtained as an ensemble average and, 
practically, can be estimated as a time-averaged quantity from a thermostatted MD or
Monte Carlo (MC) simulation at a given temperature $T_{\rm ext}$ with a given potential 
$\Phi$ for the $N$-atom system and a set of fixed reaction coordinates ${\bf X}$;
\begin{eqnarray}
\frac{\partial F({\bf X})}{\partial X_{p}} 
& = & 
\frac{1}{Z}\int d{\bf R}
\left[
      \frac{\partial \Phi({\bf R}) }{\partial X_{p}}
\right]_{\bf X} 
e^{-\Phi({\bf R})/k_{\rm B}T_{\rm ext}}
\nonumber \\ 
& \simeq & 
\frac{1}{\tau} \int^{\tau}_{0} dt
       \left[
             \frac{\partial \Phi({\bf R}(t)) }{\partial X_{p}}
       \right]_{\bf X},
\label{canonical}
\end{eqnarray}
\noindent
where 
\begin{eqnarray}
Z & = & \int d{\bf R} e^{-\Phi({\bf R})/k_{\rm B} T_{\rm ext}}.
\label{pf}
\end{eqnarray}
Here, $\tau$ is the simulation time period, and 
the $[\ \ \ ]_{\bf X}$ represents the ensemble average under 
the set of constraints.  
In the MF estimation, it is expected that the canonical 
MD or MC simulation provides 
the canonical distribution under the constraint on ${\bf X}$. 
The MF is, in our approach, evaluated using thermostatted FPMD. 
The potential energy $\Phi({\bf R})$ and 
the details of the MF evaluation will be discussed later on. 

We need to know $F({\bf X})$ to calculate the force on $X_{p}$ 
in Eq. (\ref{eqmfd}), however, $F({\bf X})$ itself is the quantity we want 
to obtain.
This problem can be solved using the conserved quantity, $H_{\rm LogMFD}$ or 
$\hat{H}_{\rm LogMFD}$ [Eq. (\ref{HMFD1}) or (\ref{HMFD2})]. 
Using this conservation law, $F({\bf X})$ can be directly evaluated 
with $\hat{H}_{\rm LogMFD}$ (when we employ a single Nos\'e-Hoover thermostat)
whose value needs to be set at the beginning of the LogMFD run;\cite{Morishita2012,Morishita2013}
\begin{eqnarray}
F({\bf X}) & = & \frac{1}{\alpha} 
                 \left[ {\rm exp} 
                    \left\{ \frac{1}{\gamma} 
                         \left(\hat{H}_{\rm LogMFD} - \sum_{p}^{L} 
                             \frac{1}{2}M_{p} \dot{X}_{p}^{2}
                         \right.
                    \right.
                 \right.
\nonumber \\ 
           &  & \left. 
                  \left. 
                    \left. 
                 - \frac{1}{2}   
\textcolor{black}{
               Q_{\eta}
}
                                  \dot {\eta}^{2} - L k_{\rm B}T_{X} \eta 
                    \right)
                  \right\}
                    - 1
                \right] .
\label{freeenergy}
\end{eqnarray}
It is required that $F({\bf X}) > 0$ at any ${\bf X}$ 
to enhance the sampling in the ${\bf X}$ subspace.
This requirement can be actually fulfilled by using appropriate values for $\hat{H}_{\rm LogMFD}$; 
$\hat{H}_{\rm LogMFD}$ should be larger than the sum of the initial kinetic energy 
for ${\bf X}$ and the initial terms for $\eta$.
See Ref. \onlinecite{Morishita2013} for details. 
Equation (\ref{freeenergy}) indicates that the $F({\bf X})$ is successively obtained along the dynamics 
of $\{X_{p}\}$, namely, ``on-the-fly".\textcolor{black}{
This means that we need not perform 
any postprocessing unlike in TI,
}
which overcomes some drawbacks of the TI 
method.
\textcolor{black}{In TI, we need to}
decompose the ${\bf X}$ subspace into many bins with a finite width, 
implying a possible missing of remarkable characters 
in the free energy due to a discretized mesh. 
In contrast, LogMFD provides $F({\bf X})$ with much higher resolution than TI, 
since LogMFD generates almost continuous {\bf X}-trajectories and 
the $F({\bf X})$ trajectories (this will be illustrated  in Fig. \ref{every}).

\begin{figure}
\includegraphics[width=5cm]{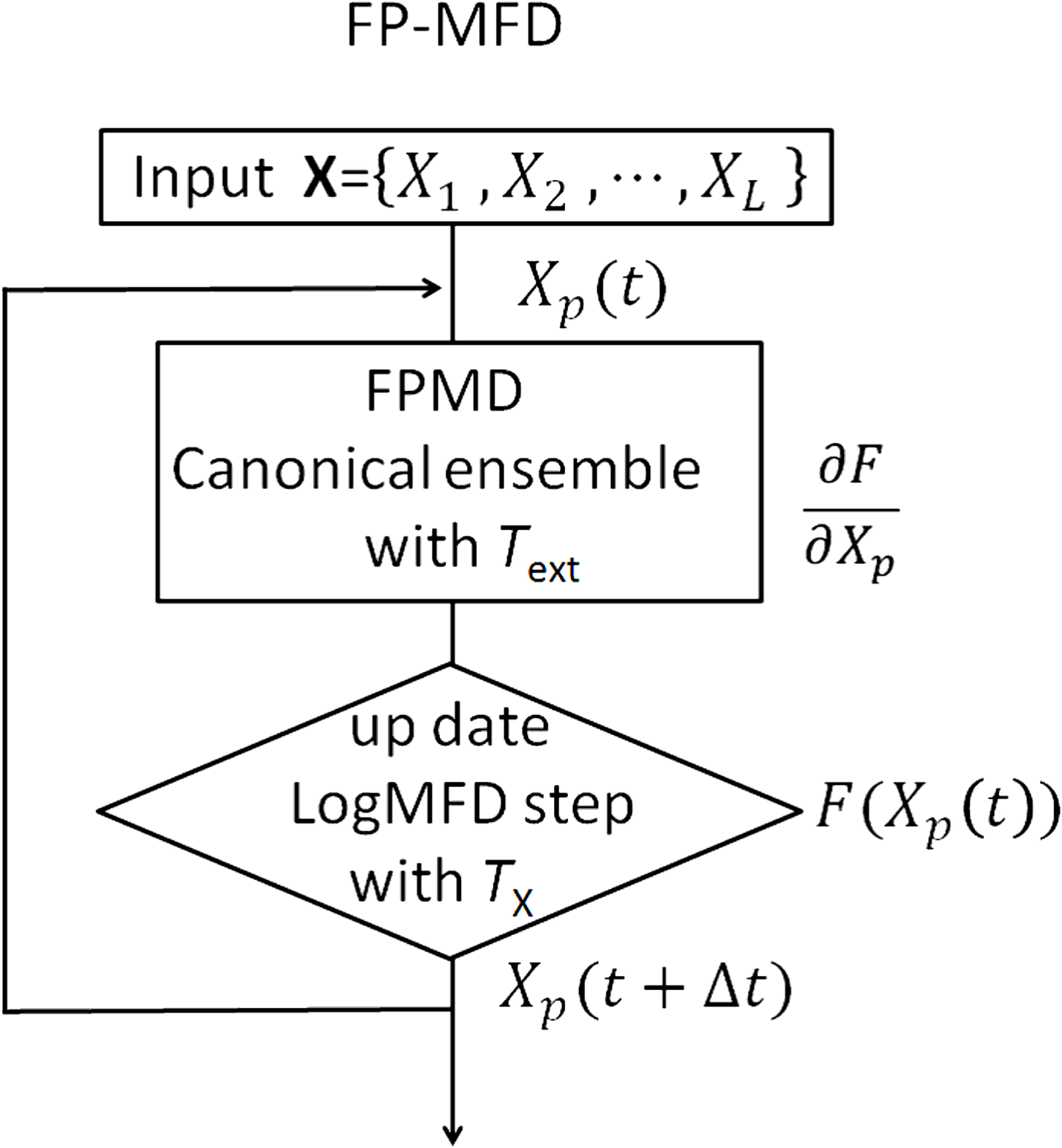}
\caption{\label{chart}
Flow chart of first-principles LogMFD.}
\end{figure}

\textcolor{black}{Summarizing LogMFD, it allows us to sample higher energy states 
efficiently and to evaluate the free energy at the local point of reaction coordinates 
without any postprocessing.}
The flow chart of the LogMFD method is displayed in Fig.~\ref{chart}. 
To update $\{ X_{p} \}$, the most important quantity is the MF, 
which is evaluated using thermostatted FPMD in our approach. 
Details of our FPMD approach are presented in the next subsection; 
Car-Parrinello molecular dynamics (CP-FPMD)\cite{Car1985}
with double Nos\'e-Hoover thermostats.\cite{Blochl1992,Morishita1999}

\subsection{\label{subsec:meanforce}First-principles mean force}

In MFD methods, the MF needs to be estimated as accurately as possible 
at the temperature $T_{\rm ext}$. The MF from first-principles 
could improve the accuracy of the free energy profiles.
We now address two technical issues associated with the evaluation
of the MF in our approach.
Firstly, the constraint on ${\bf X}$ during the FPMD run is discussed.
In order to impose a constraint on atomic coordinates, one may employ 
the SHAKE method in which a holonomic constraint is realized,\cite{SHAKE}
although complex equations should be solved for the Lagrange multiplier.   
Alternatively, the harmonic potential method can also be utilized, allowing us to 
use Eq. (\ref{canonical}) without any correction terms.
In this work, we chose the latter method.
\textcolor{black}{
Secondly,
} 
to keep a given temperature for the system and to generate the canonical 
distribution for the atomic trajectories, we employ thermostats. 

Newly developed thermostats,\cite{Morishita2010} as well as the original 
thermostat,\cite{Nose1984,Hoover1985} can also be used in conjunction with FPMD in which 
the Born-Oppenheimer(BO) surface is strictly searched 
in the time evolution\cite{Payne1992} or CP-FPMD.\cite{Car1985}

In our FPMD approach, the following energy is considered:
\begin{eqnarray}
E_{\rm tot} & = & E_{\rm BP} + E_{\rm hp},
\label{etot} \\
E_{\rm hp}  & = & \sum_{p} \frac{1}{2} k_{p} (\tilde{X}_{p}( \{ {\bf R}_{I} \} ) - X_{p})^{2}, 
\label{hp} \\
E_{\rm BP}  & = & \sum_{i} m_{\varphi}
                    \langle 
                      \dot{\varphi}_{i} | \dot{\varphi}_{i}
                    \rangle
                  + \frac{1}{2} \sum_{I} M_{I} \dot{{\bf R}}_{I}^{2}  
\nonumber \\
            &   & + \frac{1}{2} Q_{\rm e}\dot{x}_{\rm e}^{2} + 2E_{\rm kin}^{0} x_{\rm e} 
\nonumber \\
            &   & + \frac{1}{2} Q_{R}\dot{x}_{R}^{2} + g k_{\rm B}T x_{R} 
\nonumber \\
            &   & + E_{\rm fp}[\{ \varphi_{i} \},\{ {\bf R}_{I} \}],
\label{bp}
\end{eqnarray}
where $E_{\rm BP}$ and $E_{\rm hp}$ are the energy 
in the Bl\"ochl-Parrinello(BP) method\cite{Blochl1992} 
and 
the harmonic potentials for the constraint, respectively, and $E_{\rm fp}$ represents 
the potential energy in the system of electrons and ions (see Eq. (5) in Ref. 
\onlinecite{Blochl1992}). 
$x_{R}$ ($x_{\rm e}$) is the dynamical variable for the thermostat and 
$Q_{R}$ ($Q_{\rm e}$) is the corresponding mass  for $x_{R}$ ($x_{\rm e}$).
$g$ is the number of ionic degrees of freedom.  
The quantity $\tilde{X}_{p}( \{ {\bf R}_{I} \} )$ in Eq. (\ref{hp}) is
constructed from the current atomic coordinates,
which is tightly constrained to $X_{p}$ according to $E_{\rm hp}$ [Eq. (12)].
To this end, the constant $k_{p}$ is chosen 
to be a large value.  
The atomic forces come from the contributions of $E_{\rm fp}$, the thermostat, and 
the constraint $E_{\rm hp}$. These contributions result in the following equation of motion
for ${\bf R}_{I}$;
\begin{eqnarray}
M_{I}\ddot{\bf R}_{I} &=& {\bf F}_{I}^{\rm fp} - M_{I}\dot{\bf R}_{I}\dot{x}_{R}
\nonumber \\          & & - \sum_{p} k_{p} (\tilde{X}_{p}( \{ {\bf R}_{I} \} )-X_{p}) 
                              \frac{\partial \tilde{X}_{p}( \{ {\bf R}_{I} \} )}
                                 {\partial {\bf R}_{I}}.  
\label{cp4}
\end{eqnarray}
The equations of motion for the wavefunction ($\varphi_{i}$) and the 
heat baths ($x_{\rm e}$ and $x_{R}$) are not changed from the original BP method
by introducing the constraint, implying a less effort 
for converting a conventional computational code to the present one. 

According to Eq. (\ref{canonical}), the first-principles mean force is obtained 
as a time average of 
$-\partial E_{\rm tot}/\partial X_{p}$:
\begin{eqnarray}
-\frac{\partial F({\bf X})}{\partial X_{p}} & \sim &  k_{p} 
                          \langle 
                                 \tilde{X}_{p}( \{ {\bf R}_{I} \} )-X_{p} 
                          \rangle , 
\label{mf}
\end{eqnarray}
where $\langle \ \ \ \rangle$ represents a canonical ensemble or a time average. 
This formula is general as far as the atomic configuration samples the canonical distribution
under the required constraint. This implies that 
the relation of Eq. (\ref{mf}) is also useful in the TI method, as explained in 
Appendix A. 
In order to show an achievement of the constraint and the temperature control, 
we present typical time evolution of the reaction coordinate 
$X_{p}$ in the next section. 
  
In CP-FPMD, the fictitious kinetic energy of the wave function, 
namely, the first term in Eq. (\ref{bp}), should follow the dynamics of  $\{ {\bf R}_{I} \}$
as quickly as possible.\cite{Blochl1992} 
To this end, it is important to find an appropriate $E_{\rm kin}^{0}$. 
For a given atomic configuration, (i) we start a CP-FPMD run with the system exactly 
on the BO surface. (We converge the electronic state to the BO surface beforehand.) 
Then (ii) we perform the CP-FPMD run for a few tens of MD steps without the heat baths 
(may be better, without the constraint). 
During this period, the system slightly leaves the exact BO surface.  
(iii) If the temperature of the system reaches $T_{\rm st}$
within the given period, then we set the value of the kinetic energy of 
the wave functions at the moment as $E_{\rm kin}^{0}$. Due to a practical reason for stabilizing 
the simulation, $T_{\rm st}$ is taken as $\sim 0.85 T_{\rm ext}$.
(iv) When the system does not reach an appropriate temperature, we introduce a new atomic 
configuration by distorting the previous atomic configuration. 
We restart the process from (i). 
After setting $E_{\rm kin}^{0}$, we switch on the thermostats in 
the BP method accompanied with the constraint of Eq. (\ref{hp}). 
The rest of the FPMD steps are used for the MF evaluation of Eq. (\ref{mf}). 
In the computation mentioned above, the initial atomic configurations for each of the series
of the CP-FPMD runs were taken from 
the atomic configuration in the CP-FPMD runs previously done.
We will detail the procedure to perform FP-LogMFD calculations,
including parameter settings, in the next section.

\begin{figure}
\includegraphics[width=6cm]{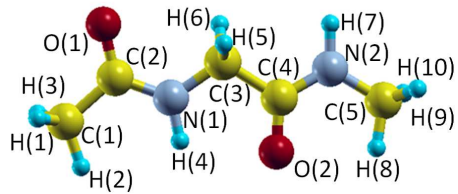}
\caption{\label{glycine} 
Atomic structure of the glycine dipeptide molecule 
with atomic specification. 
Two dihedral angles,
$\phi$ and $\psi$, are formed by the atomic series C(2)-N(1)-C(3)-C(4) 
and N(1)-C(3)-C(4)-N(2), respectively. 
The figure displays the atomic configuration with
\textcolor{black}{
$(\phi, \psi)=(180^\circ , 180^\circ)$}
.}
\end{figure}

\section{\label{sec:demonstration}NUMERICAL DEMONSTRATION}

\subsection{\label{subsec:molecule}Molecular configuration}

To illustrate our {\it ab initio} approach to free energy reconstruction, we consider the free
energy profile of glycine dipeptide molecule 
[2-(Acetylamino)-N-methylacetamide] in vacuum, 
as shown in Fig.~\ref{glycine}. The atoms are specified by 
the symbol with numbering of C(1), C(2), $\cdots$ , O(1), O(2), .. etc.. 
from the left-hand-side of the figure. 
The two dihedral angles are labeled as $\phi$ and $\psi$, which are formed by the atomic 
series C(2)-N(1)-C(3)-C(4) and N(1)-C(3)-C(4)-N(2), respectively. 
In other words, these angles are formed by the plane of N(1)-C(3)-C(4) and the plane 
associated with the peptide bond (-OCNH-). In nature, the latter plane in proteins 
has usually observed as the {\it trans}-form rather than 
as the {\it cis}-form.\cite{text-book} 
Actually, in our calculation, the {\it cis}-form of the right-hand-side of the peptide bonds 
in Fig.~\ref{glycine} is higher in energy by 2.1 kcal/mol than the form
presented in Fig. \ref{glycine}. 

In this section, we demonstrate the application of FP-LogMFD to the 
glycine dipeptide molecule. 
We have obtained the free energy landscape $F(\phi, \psi)$ with respect to 
the dihedral angles $\phi$ and $\psi$ at room temperature 300 K ($=T_{\rm ext}$).
First, FP-LogMFD with the fixed dihedral angle $\phi$ was
performed to set the parameters required and to obtain the one-dimensional
free energy profile along $\psi$. 
Then, the FP-LogMFD runs in the $\phi$-$\psi$ space were performed, revealing
the details of the two-dimensional free energy landscape.
We have also performed TI calculation to reconstruct the one-dimensional
free energy profile, which is compared to the FP-LogMFD result for benchmarking.

\subsection{\label{subsec:parameter}Parameter setting}

\begin{figure}
\includegraphics[width=7.5cm]{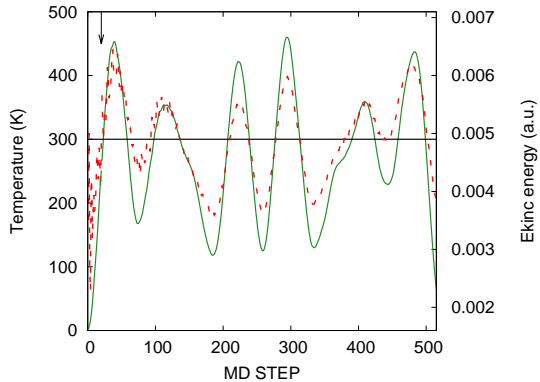}
\caption{\label{kineticenergy} 
Time evolution of the instantaneous temperature (green full curve, left scale) and 
the kinetic energy of the wave functions (red dashed curve, right scale) in the FPMD run
at $T_{\rm ext}=300$ K with the constraints
\textcolor{black}{ 
$\phi=-80.0^\circ$ and $\psi=-76.8^\circ$. 
}
The horizontal line indicates the values of
$T_{\rm ext}$ and $E_{\rm kin}^{0}$.
The arrow indicates the time step at which the 
constraint on the dihedral angles and the temperature control is turned on.}
\end{figure}

For the CP-FPMD runs, we have used the plane wave basis set and
density functional theory with the generalized gradient 
approximation(GGA).\cite{KS,Perdew92}
The energy cutoffs of 25 and 250 Ry are taken for electronic wave function and charge density, 
respectively.\cite{Pasquarello1992} The ultrasoft pseudopotentials are 
used.\cite{Vanderbilt1990} The $\Gamma$-point sampling is adopted for the molecular 
system placed in a cubic box with the dimension of 20 a.u.(10.58 \AA). 
For the canonical FPMD simulation in the framework of the BP method, 
the time step is set to 10 a.u. ($\sim$ 0.24 fs). 
This is a typical value for the CP method.\cite{Car1985}   
The parameters for $Q_{R}$, $Q_{\rm e}$, and $m_{\varphi}$ are set to 
$5 \times 10^{5}$ a.u.,
$5 \times 10^{3}$ a.u.,\textcolor{black}{\cite{Oda2002}} and 200 a.u.,\textcolor{black}{\cite{Oda2004}} respectively.  
$E_{\rm kin}^{0}$ in Eq. (\ref{bp}) is automatically determined by the anzatz 
described before (see Sec. \ref{subsec:meanforce}).
Figure \ref{kineticenergy} presents the time evolution of the kinetic energy 
of the wave functions and the instantaneous temperature 
of the molecular system (proportional to the kinetic energy of 
atoms). 
In the present calculations, $E_{\rm kin}^{0}$ was set to be 0.0049 a.u.

\begin{figure}
\includegraphics[width=7.5cm]{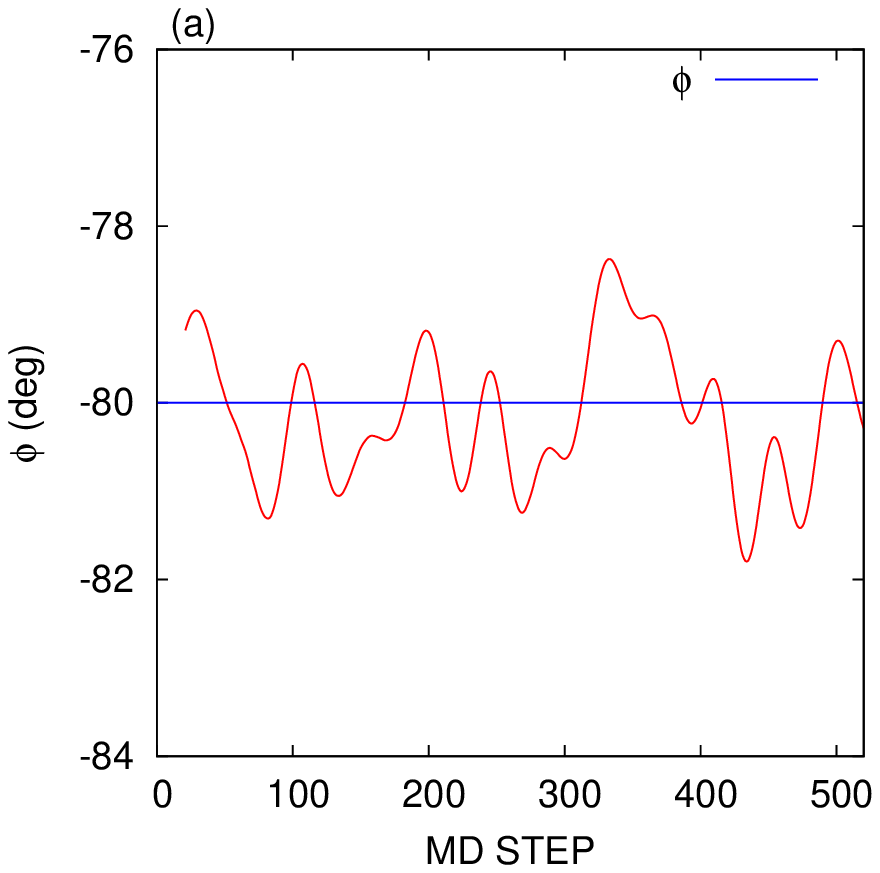}
\includegraphics[width=7.5cm]{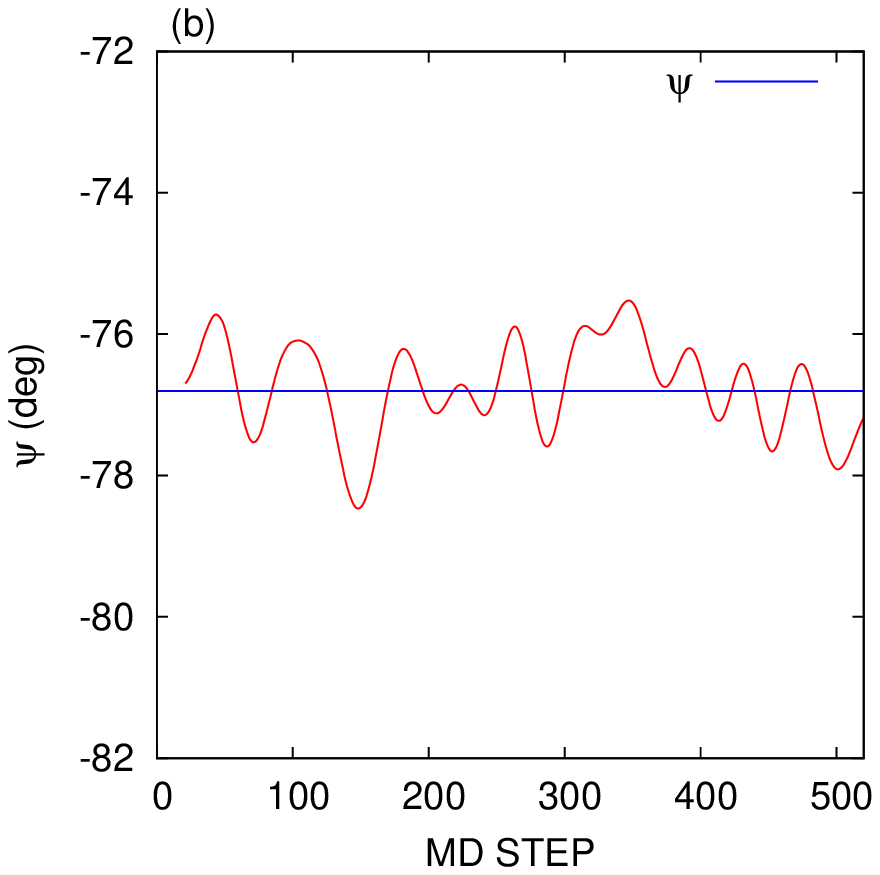}
\caption{\label{dihed}
Time evolution of 
the instantaneous dihedral angles, 
(a) $\tilde{\phi}$ and  (b) $\tilde{\psi}$, 
in the FPMD run
with the constraints, 
\textcolor{black}{
$\phi=-80.0^\circ$ and $\psi=-76.8^\circ$}
(horizontal lines).}
\end{figure}

In this demonstration, the following harmonic potentials are employed to constraint $\tilde{\phi}$ and $\tilde{\psi}$;
\begin{eqnarray}
E_{\rm hp}  & = &  \frac{1}{2} k_{\phi} 
                    (\tilde{\phi}( \{ {\bf R}_{I} \} ) - \phi)^{2}
                  +\frac{1}{2} k_{\psi} 
                    (\tilde{\psi}( \{ {\bf R}_{I} \} ) - \psi)^{2}, 
\label{harmonicpotential}
\end{eqnarray}
where $\phi$ and $\psi$ are the target dihedral angles [$X_{p}$ in Eq. (\ref{hp})]
and $\tilde{\phi}$ and $\tilde{\psi}$ are the temporal ones determined 
from the instantaneous molecular configuration [$\tilde{X}_{p}$ in Eq. (\ref{hp})]. 
Both of $k_{\phi}$ and $k_{\psi}$ are taken to be 2.4 a.u./${\rm rad}^{2}$ 
(\textcolor{black}{0.46 kcal/mol/${\rm deg}^{2}$}). 
Figure \ref{dihed} shows typical time evolution of $\tilde{\phi}$ 
and $\tilde{\psi}$ 
with \textcolor{black}{$(\phi, \psi)=(-80.0^\circ, -76.8^\circ)$}.
This figure indicates that the temporal $\tilde{\phi}$ and $\tilde{\psi}$ 
fluctuate around the respective given value, implying the constraint 
to be imposed correctly.

\begin{figure}
\includegraphics[width=7.5cm]{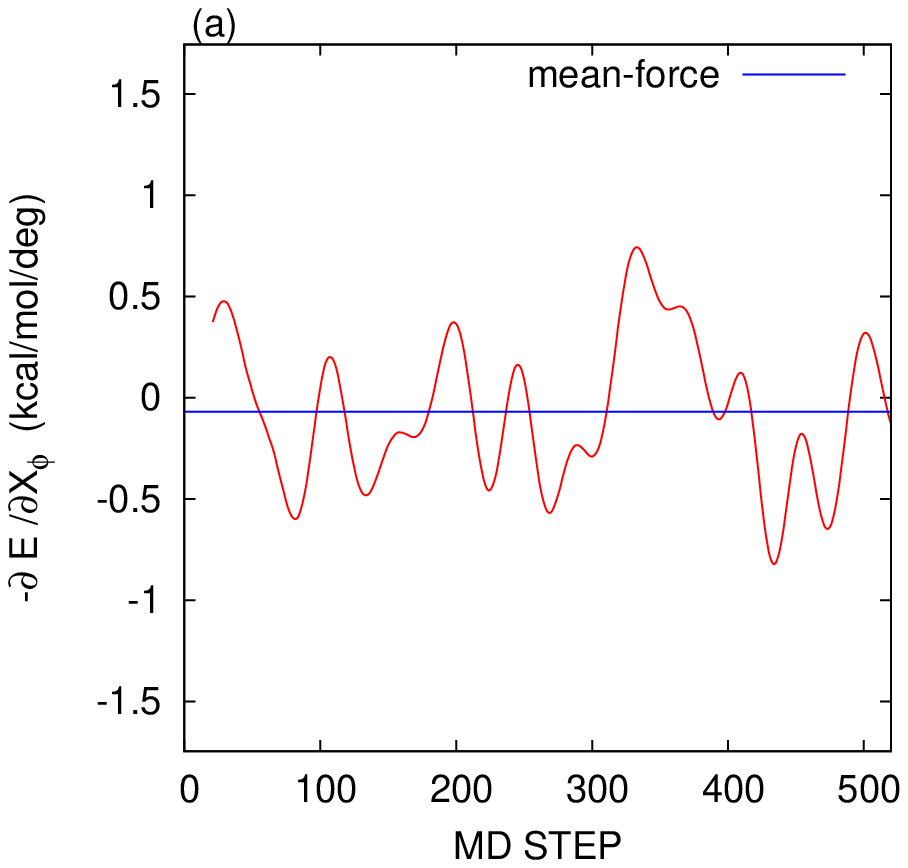}
\includegraphics[width=7.5cm]{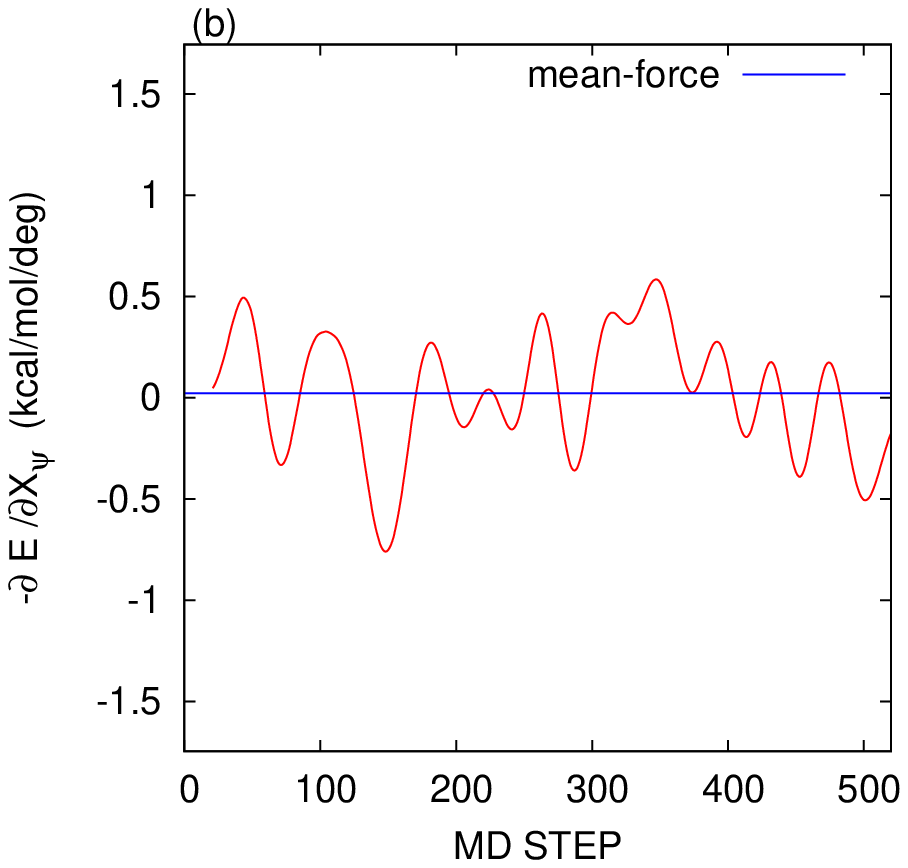}
\caption{\label{meanforce} 
Time evolution of the instantaneous force, 
(a) $k_{\phi}(\tilde{\phi}-\phi)$ and (b) $k_{\psi}(\tilde{\psi}-\psi)$, 
accompanied with the horizontal lines which indicates 
the mean forces, $-\partial F/\partial \phi$ 
and $-\partial F/\partial \psi$.}
\end{figure}

The time evolution of the temporal force is presented in Fig.~\ref{meanforce}.
Averaging over 500 steps (from the 21th step to 520th step), 
the mean forces acting on $\phi$ and $\psi$ were estimated to be
\textcolor{black}{ 
$-$0.06863 and 0.02220 (kcal/mol)/deg}, 
respectively
(the fluctuations are 
limited to 
\textcolor{black}{
$\pm 0.87$ (kcal/mol)/deg
}
).
The accuracy of 
the MF strongly depends on the number of MD steps, defined as $N_{\rm BP}$. 
In fact, we found that the decrease of $N_{\rm BP}$ (from 500 steps to 300 steps) 
deteriorated the MF, and the resultant free energy profile became much worse, 
compared to those by $N_{\rm BP}=500$.
$N_{\rm BP}$ was thus set to 500 steps 
in the present study. 

Evaluation of the MF is also needed in the TI method. 
The MF at each of the grid points in the TI calculation was obtained by 
averaging the instantaneous forces from a set of many FPMD runs with 
Eqs. (\ref{etot}) and (\ref{hp}). 
The successive simulation started with random atomic distortions from the 
previous atomic configuration. In the present study, we have performed
120 FPMD runs, each consisting of 600 FPMD steps, i.e., 72,000 FPMD steps in total
for each grid point of the reaction coordinate.
60,000 FPMD steps out of the 72,000 steps were devoted to estimation of the MF
at a single grid point.

For a set of given coordinates $(\phi, \psi)$, as shown in Fig.~\ref{chart}, 
$-\partial F/\partial \phi$ and $-\partial F/\partial \psi$ 
were estimated for the hypothetical dynamics given 
by Eqs. (\ref{eqmfd}) and (\ref{eqmfdeta}) with $T_{X}=300$ K in FP-LogMFD. 
A single Nos\'e-Hoover thermostat\cite{Nose1984,Morishita2010,Hoover1985}
was used.
In the FP-LogMFD runs, 
the variables of $\phi$, $\psi$, and $\eta$ were updated 
using a time step of 1 $\tau$, with the masses of \textcolor{black}{$M_{\phi(\psi)}=1.7\times 10^{4}$ (kcal/mol)/(deg/$\tau^{2}$)}
and $Q_{\eta}=L k_{\rm B}T_{\rm X} \tau_{\eta}^{2}$ with $\tau_{\eta}=50$ $\tau$, 
where $\tau$ represents the time unit.
[the time unit can, in fact, be arbitrarily chosen, e.g., $\tau$=1 fs, 
since the dynamics of $\phi$ has nothing to do with the resultant
$F(\phi)$.]
The parameters of $\alpha$ and $\gamma$, which determines the degree
of the effective reduction of the free energy
barriers, were taken as $\alpha=3$ (kcal/mol)$^{-1}$ and $\gamma=1/\alpha$, 
with this value of $\gamma$ corresponding to 170 K. 
After solving Eqs. (\ref{eqmfd}) and (\ref{eqmfdeta}), 
the conversion to $F(\phi, \psi)$ was performed 
using Eq. (\ref{freeenergy}) with 
$\hat{H}_{\rm LogMFD}=1$ kcal/mol. 
$\hat{H}_{\rm LogMFD}$ should be 
set to ensure $\alpha F_{\rm min}(\phi, \psi) + 1 > 0$,
where $F_{\rm min}$ is the minimum of the free energy. 
Note however that there is, in principle, no upper limit for 
the value of $\hat{H}_{\rm LogMFD}$.\cite{Morishita2013}

The validity of the LogMFD results mainly depends 
on the accuracy of the MF, 
which influences the conservation of $\hat{H}_{\rm LogMFD}$ [Eq. (\ref{HMFD2})]. 
As was already mentioned, the quality of the MF 
can be controlled by $N_{\rm BP}$ and 
the mass parameter $M_{\phi(\psi)}$.\cite{Morishita2013}
The increase of 
$M_{\phi(\psi)}$, which reduces (suppresses) the velocity of
the dynamical variables, 
results in a more accurate profile for the MF, and thus, the free energy profile. 
We found, by decreasing the $M_{\phi(\psi)}$ by the factor ten, 
that the difference between the LogMFD and TI results becomes
0.22 kcal/mol from 0.18 kcal/mol on average.
In the present system, the periodicity with respect to $\phi$ and $\psi$ can be available 
for checking the accuracy of simulations.

\subsection{\label{subsec:1d}One dimensional profile}

\begin{figure}
\includegraphics[width=7.5cm]{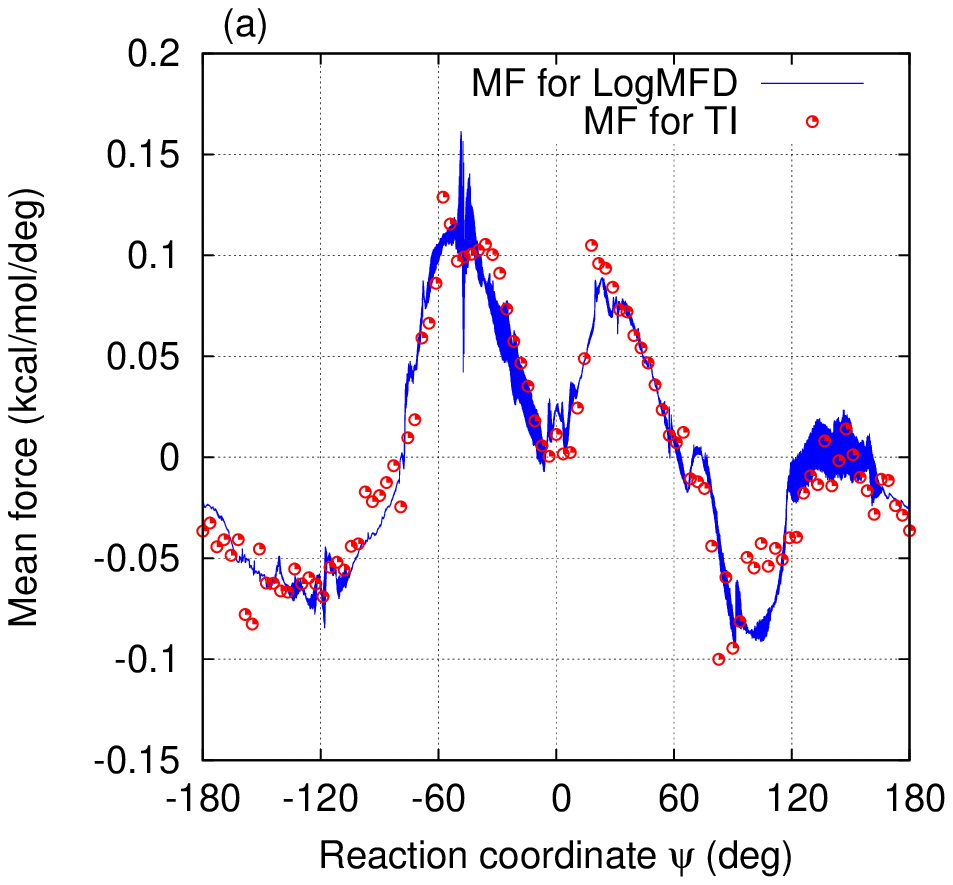}
\includegraphics[width=7.5cm]{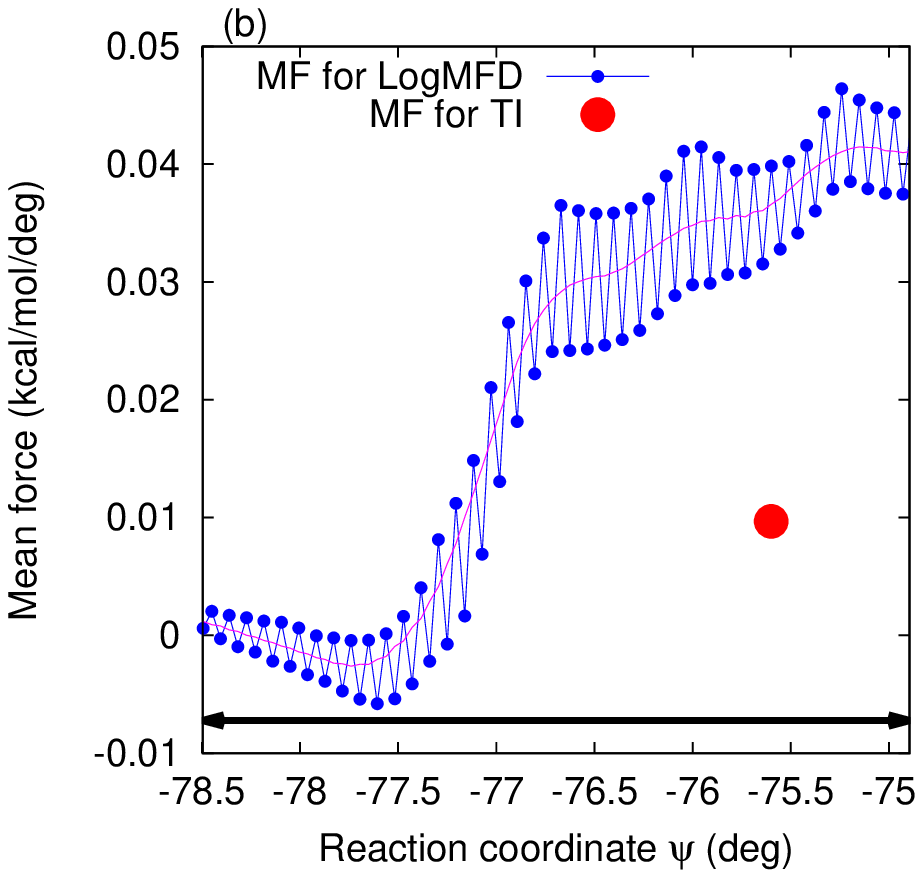}
\caption{\label{every}
(a) MF
with respect to the dihedral angle $\psi$ 
in the glycine dipeptide molecule, constraining the other dihedral 
angle $\phi$ to 
\textcolor{black}{
$-$80$^{\circ}$
}.
The blue curve and red symbols indicate the MF calculated
from the LogMFD and TI calculations, respectively.
(b) The magnified profile of the MF (vibrational curve) 
around 
\textcolor{black}{
$\psi=-76.8^\circ$}, 
with a smooth curve showing
the profile obtained 
by averaging over ten MFD time steps.   
The arrow indicates the width of the mesh used in the 
TI calculation,
showing that the MF around this $\psi$ range is approximated 
by only a single grid-point result.
}
\end{figure}

\begin{figure}
\includegraphics[width=7.5cm]{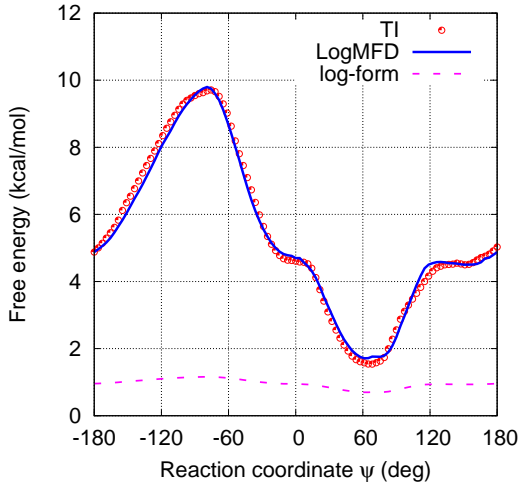}
\caption{\label{logmfd-ti}
Free energy profiles
with respect to the dihedral angle $\psi$ in the glycine 
dipeptide molecule, constraining the other dihedral angle 
$\phi$ to 
\textcolor{black}{
$-$80$^{\circ}$}, 
obtained from
the LogMFD (blue curve) and TI
(red dots) calculations.
The logarithmic energy ($\gamma \log (\alpha F(\psi)+1$) )
(magenta curve) is also presented for comparison, indicating 
a substantial reduction of the free energy barrier.}
\end{figure}

For demonstrating the free energy evaluation using FP-LogMFD, 
we performed FP-LogMFD simulations for the dynamical 
variable $\psi$ while 
keeping $\phi$ to be 
\textcolor{black}{$-$80$^{\circ}$}.
In Fig.~\ref{every}(a), the MF profiles from the LogMFD and 
TI calculations are presented, showing the LogMFD result
is in good agreement with the TI result.
Figure \ref{every}(a) also shows that there are regions where 
the MF drastically varies 
in a narrow range, e.g., 
\textcolor{black}{
$\psi = -115^\circ \sim -57.3^\circ$}.
In Fig.~\ref{every}(b), the magnified profile in the range of 
\textcolor{black}{
$-78.5^\circ \leq \psi \leq -74.9^\circ$}
indicates 
that, although the data by LogMFD shows a 
vibrational behavior, 
the MF averaged over 10 MFD steps varies smoothly. 
This behavior of the MF in LogMFD is remarkable
when the profile 
exhibits a rapid variation. 
As shown in Fig.~\ref{every}(b),  
a set of uniformly sparse grid points is only used
in the TI method due to a limited computational resources. 
LogMFD thus can provide missing data in between
each of the grid points
in the TI calculations without much additional 
computational cost. 

Figure~\ref{logmfd-ti} shows the free energy profiles 
obtained by the LogMFD and TI methods. 
Each of the free energy profiles is shifted to have the same value
(5 kcal/mol) at 
\textcolor{black}{
$\psi=-180^\circ$} 
for comparison in Fig. ~\ref{logmfd-ti}.
LogMFD runs were initiated at 
\textcolor{black}{
$\psi=57.3^\circ$} 
(around the minimum) to either direction (with increasing or 
decreasing $\psi$) with $T_{\rm X}=300$ K and 
were ended at 
\textcolor{black}{
$\psi=92^\circ$ 
}
after passing through the periodic
boundary at 
\textcolor{black}{
$180^\circ$ or $-180^\circ$}. 
It should be remarked that the value of 
\textcolor{black}{
$F(\psi=92^\circ)$} 
estimated when
\textcolor{black}{
$\psi=92^\circ$} 
was sampled for the first time is almost the same
as the 
\textcolor{black}{
$F(\psi=92^\circ)$ 
}
estimated when 
\textcolor{black}{
$\psi=92^\circ$} was sampled the second time,
indicating the energy dissipation, which degrades the accuracy of $F(\psi)$, is negligible.

There is the maximum at 
\textcolor{black}{
$\psi=-77.9^\circ$}, 
and the minimum at
\textcolor{black}{ 
$\psi=63.0^\circ$} 
in the profile, 
as shown in Fig.~\ref{logmfd-ti}. 
We stress here that the
dynamics for the reaction coordinate $\psi$ was very smooth,
even the large energy barrier exists.
The difference between the minimum and maximum free 
energy approximately amounts to 8 kcal/mol, 
corresponding to about 4024 K. 
This energy difference was entirely suppressed by 
the logarithmic form.
Figure~\ref{logmfd-ti} also shows the effective potential curve
of $\gamma \log ( \alpha F(\psi) + 1 )$, 
indicating that the actual energy barrier for $\psi$ 
became $\sim$ 0.8 kcal/mol, comparable to 402 K.  
Such a substantial reduction of the energy barrier can be 
controlled by the parameters ($\alpha$ and $\gamma$).

\begin{figure}
\includegraphics[width=7.5cm]{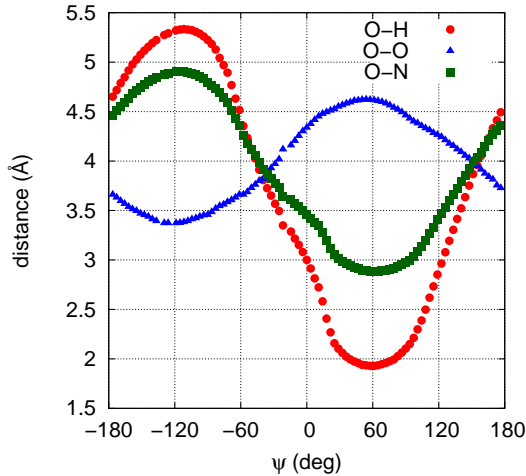}
\caption{\label{distance}
Atomic distances for O(1)-H(7), O(1)-O(2), and O(1)-N(2) 
as a function of $\psi$.}
\end{figure}

Before proceeding to the two dimensional landscape, we discuss 
the one-dimensional free energy profile in more detail. 
As pointed out, there are the minimum and maximum  
in the profile.
The former and latter are related to a hydrogen bond and 
a rendezvous of a pair of the oxygen atoms in the peptide bonds, respectively.  
Other characteristic properties are found around 
\textcolor{black}{
$\psi=0^\circ$ and $140^\circ$}, 
where the free energy shows 
\textcolor{black}{
a profile with zero curvature}.  
We consider that this is due to
breaking of the hydrogen bond which is formed 
around 
\textcolor{black}{
$\psi=57.3^\circ$}. 
This consideration is supported by the fact that 
the MF in the corresponding part 
\textcolor{black}{is} 
almost zero 
(see Fig.~\ref{every}). 
Atomic distances as a function of $\psi$ are shown
in Fig.~\ref{distance}. From this figure, 
the free energy minimum in Fig.~\ref{logmfd-ti}
is found to appear around the minimum distance of O(1)-H(7) and 
O(1)-N(2), while the energy maximum appears around the minimum 
distance of O(1)-O(2). The latter case may correspond to a 
large electric dipole state for the molecule.
The distance of 2.5 $\sim$ 3 \AA\ for O(1)-H(7) at
\textcolor{black}{
$\psi=0^\circ$ and $120^\circ$}
is out of the range of the hydrogen bonding, 
where the MF is $\sim$ 0. 
From this, we consider
that an energy of about 3 kcal/mol is gained by the hydrogen bond 
(see Fig.~\ref{logmfd-ti}). 
This energy is comparable to a typical bonding energy of the hydrogen 
bond (3 $\sim$ 10 kcal/mol) reported
in a literature.\cite{text-book}


\subsection{\label{subsec:2d}Two dimensional profile}

\begin{figure}
\includegraphics[width=8.0cm]{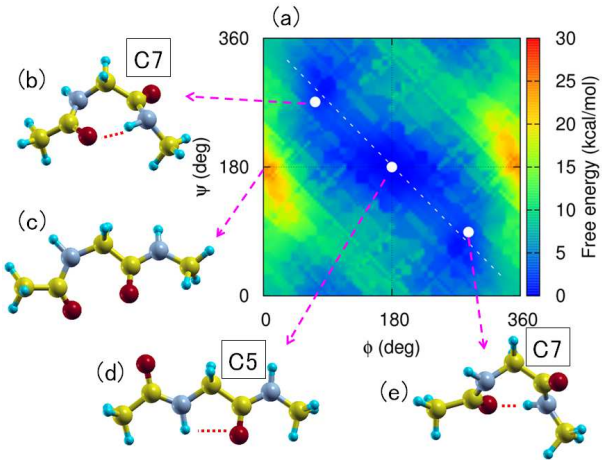}
\caption{\label{freemap2}
(a) Free energy contour map $F(\phi,\psi)$  
and (b)--(e) typical atomic configurations at three stable 
states (C5 and C7 atomic configurations) and an unstable state.
The white bullets indicate the positions for the stable states.
A pass way that approximately connects these stable states
with a straight line is displayed in a white dashed line
(see also Fig. \ref{feC5C7}).
}
\end{figure}

\begin{figure}
\includegraphics[width=7.5cm]{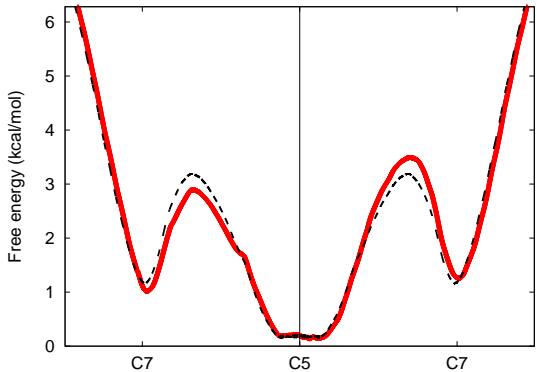}
\caption{\label{feC5C7}
The one dimensional free energy profile along the line 
that approximately connects the C5 and C7 states.
The full and dotted curves represents the bare and symmetrized plot of 
$F(\phi,\psi)$.}
\end{figure} 

For constructing the two dimensional free energy profile, the dynamical
equations for both of
$\phi$ and $\psi$ were used. 
Supposing 
that the free energy minimum in the $\phi$-$\psi$ space may be lower than
that in the one-dimensional $\psi$ space at 
\textcolor{black}{
$\phi=-80^\circ$},
$\hat{H}_{\rm LogMFD}$ was 
increased to 1.2 kcal/mol to 
shift the baseline of the free energy landscape.
The temperature $T_{X}$ and $\alpha$ were chosen to be
\textcolor{black}{
small values};
$T_{X}=200$ K and $\alpha$ =2 (kcal/mol)$^{-1}$
\textcolor{black}{
to suppress numerical errors in longer simulations},
while 
the other parameters took the same values as used 
in the one dimensional FP-LogMFD calculations. 
The two-dimensional FP-LogMFD runs were started from the minimum of the one dimensional 
profile of $F(\psi)$
and were extended to four directions. 
The branch off can be performed from one simulation to others. 
These simulations can be performed independently, implying 
that parallel treatment is highly effective in LogMFD.\cite{Morishita2013}

Figure~\ref{freemap2} shows the free energy contour map 
in the $\phi$-$\psi$ plane with typical molecular configurations. 
The glycine dipeptide molecule has an intrinsic mirror symmetry in its 
atomic geometry. Atomic structures which are related to each other
by the mirror operation with respect to the N(1)-C(3)-C(4) plane 
has the same energy in gas phase. This feature should also be seen in 
the free energy landscape $F(\phi, \psi)$. 
Therefore, the statistical errors can be reduced by symmetrizing 
the two-dimensional free energy  
with respect to the point of 
\textcolor{black}{
$(\phi, \psi)=(180^\circ, 180^\circ)$} 
(there is the inversion symmetry in the map). 
A non-symmetrized free energy profile
along the white dashed line in Fig. \ref{freemap2}
is presented in the last paragraph in this subsection.

The free energy landscape (Fig. \ref{freemap2}) shows that there are three stable states 
(three energy valleys) and a series of unstable states (energy mountains).  
The most stable state appears around
\textcolor{black}{
$(\phi, \psi)=(180^\circ, 180^\circ)$} ,
whose atomic configuration is presented in Fig.~\ref{glycine}
(or Fig.~\ref{freemap2}(d)). 
This is assigned to the C5 configuration\cite{Cheam1989} and 
is stabilized by the hydrogen bond, the five-membered ring, 
and the configuration with separated oxygen atoms (almost zero electric dipole). 
The other two stable states, found around 
\textcolor{black}{
$(\phi, \psi)=(288^\circ, 88^\circ), 
(72^\circ, 272^\circ)$}, 
are assigned to the C7 configuration and are also 
stabilized with the hydrogen bond, the seven-membered ring, 
and the configuration with moderately separated oxygen atoms 
(small electric dipole). 
The free energy for the C7 configuration is higher by 0.58 kcal/mol than that 
for the C5 configuration. 
This energy difference is quite small and comparable to 290 K. 
The total (internal) energy computation also indicates that the C5 
configuration is either lower in energy than the C7 by 0.38 kcal/mol, while 
the work by the quantum chemistry calculation reports that 
the C5 is \textcolor{black}{either lower than the C7 by 0.58 kcal/mol,\cite{Quentin1992} or higher by 1
and 0.58 kcal/mol.\cite{Fujitani2009,Klimkowski1985} 
}
The atomic configuration of the most unstable state is presented 
in Fig.~\ref{freemap2}(b).
This instability comes from an assemble of oxygen atoms in the molecule
(implying a large electric dipole). 
The energy barrier measured from the bottom of the free energy landscape
(highest energy mountain)
amounts to 26 kcal/mol, corresponding to 
13000 K and to 730 K with
$ \gamma {\rm log} (\alpha F+1)$. Again, 
LogMFD enables to sample such higher energy configuration 
in the same footing used around the ground state. 

It is interesting to see the transition from the most stable state 
to another stable state. The one dimensional free energy profile 
roughly linking the C7 and C5 configurations
is presented in Fig.~\ref{feC5C7}, as a typical energy profile. 
For simplicity,
the pass way of reaction coordinate was assumed to 
be along the straight line which connects the two states near 
the C5 and C7 states in the $\phi$-$\psi$ plane, as specified 
in Fig.~\ref{freemap2}. 
From Fig. \ref{feC5C7}, the energy barrier between C5 and C7 
configurations is estimated to be about 3 kcal/mol 
when measured from the C5 configuration. 
The energy differences between the C5 and C7 states shown in Fig.~\ref{feC5C7} 
are about 1 kcal/mol. These values are slightly larger 
than the value reported above (0.58 kcal/mol)
because of the approximate pass way (this approximation causes the uncertainty
of about 0.4 kcal/mol).

\subsection{\label{subsec:efficiency}Computational efficiency}

In constructing the free energy profile (Fig. \ref{logmfd-ti}), 
4 $\times$ 10$^{6}$ FPMD steps were devoted in 
the FP-LogMFD calculation,  
while 7.2 $\times$ 10$^{6}$ FPMD steps were needed in the TI calculation. 
About 45 \% of the computational cost was saved. 
This demonstrates a good efficiency of LogMFD 
in the computational cost. 
In addition, in the course of the construction of the two dimensional 
profile (Fig.~\ref{freemap2}), we carried out a set of LogMFD runs which,
in total, sampled
1.2 $\times$ 10$^{8}$ FPMD steps (configurations). 
Even though the accuracy in the two dimensional profile may be slightly reduced,
the computational cost is only 30 times larger than 
that in the one dimensional calculation.

\section{\label{sec:results}DISCUSSIONS}

\begin{figure}
\includegraphics[width=7.5cm]{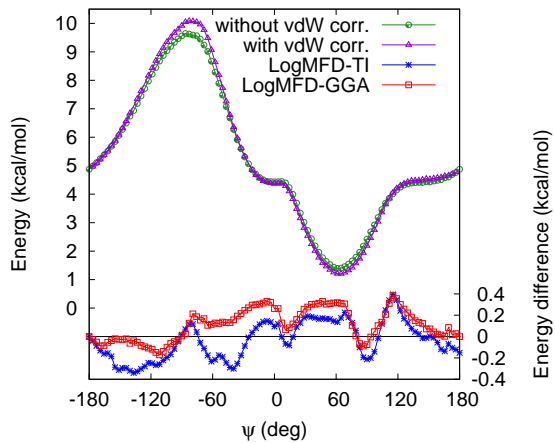}
\caption{\label{dft}
Total (internal) energy 
for the glycine dipeptide molecule as a function 
of $\psi$ keeping 
\textcolor{black}{
$\phi=-80^{\circ}$},
which was
obtained from the density functional theory (DFT) 
calculation
with (purple triangle symbols) or without (green circle symbols) 
the van der Waals(vdW) correction.
The blue asterisks denote the difference between the free energy
obtained by LogMFD and that by TI, while the red squares
denote the difference between the free energy
by LogMFD and the internal energy without the vdW correction.
}
\end{figure}
As mentioned in Sec. \ref{subsec:molecule}, the peptide bond takes 
the {\it trans}- or {\it cis}-form. In our simulations, 
the {\it trans}-form has been entirely observed and the statistical sampling 
of the {\it cis}-form has been missed. This is because the barrier between 
the {\it trans}- and {\it cis}-forms may be extremely high,
and also because the reaction coordinates chosen in the present LogMFD
calculations may not be suitable for sampling the {\it cis}-form.
If one needs to sample the {\it cis}-form, incorporation of additional 
reaction coordinates is of use, which is easily realized in LogMFD. 

It is interesting to see the contribution of the entropy in the free energy. 
We have calculated the total energy (the internal energy) 
as a function of $\psi$ keeping 
\textcolor{black}{
$\phi=-80^{\circ}$}, 
as shown in Fig.~\ref{dft}.  
The grid points used in the calculation of the internal energy
are the same as used in the TI calculation. 
The internal energy profile is very similar to the free energy profile 
(Fig.~\ref{logmfd-ti}). 
The difference between the free energy and the internal energy
is found to be within 0.5 kcal/mol
(if the energy scale is adjusted to give zero entropy at 
\textcolor{black}{
$\psi=-180^\circ$}), implying a small contribution from the entropy.
We roughly estimated the uncertainty of the free energy as $\sim$ 0.4 kcal/mol,
which is comparable to the variation of the entropy with $\psi$.
It is thus considered that the entropic contribution is hardly changed
with $\psi$.
This is not surprising because 
the number
of possible conformations in the present system is relatively small, which
does not significantly depend on the dihedral angles.
Also, the glycine dipeptide molecule is in vacuum, not in a solvent.
We however stress that LogMFD is able to unveil the variation of the entropy, if any,
which is, for example, seen in our preliminary calculations for a model system
of protein-G consisting of 56 amino acids.\cite{Isobe2001} 

The free energy profile for the glycine dipeptide molecule
was previously obtained using \textcolor{black}{classical LogMFD with} an empirical 
force field.\cite{Morishita2012,Morishita2013}
The profile is similar to that obtained using FP-LogMFD in this work, 
indicating the validity of the empirical force-field to some extent.
There are, however, some differences in the profile. 
As pointed out in Sec. \ref{subsec:1d}, we observe the 
\textcolor{black}{
zero curvature}
around
\textcolor{black}{
$\psi=0^\circ$ and $140^\circ$}.
This behavior is also seen in the internal energy profiles  
(Fig. \ref{dft}) in the FP-LogMFD approach.  
In fact, the explicit inclusion of the van der Waals
interaction\cite{Dion2004,Cooper2010,Obata2013}
into 
\textcolor{black}{
the DFT(GGA)}
calculations
does not change the overall profile of the internal energy
\textcolor{black}{
(note that the binding energy is underestimated using GGA)}. 
It is thus considered that the 
\textcolor{black}{
zero curvature} 
is not attributed to an inappropriate DFT description,
while the linear behavior observed around 
\textcolor{black}{
$\psi=0^\circ$ and $140^\circ$}
in the previous results
may come from insufficient transferability of the empirical force field.

\section{\label{sec:summary}SUMMARY}

We have demonstrated that the {\it ab initio} based MF
can be incorporated 
into the LogMFD method, which improves the reliability 
and accuracy in the free energy calculation. 
FP-LogMFD has been applied to reconstruction
of the free energy landscapes of the glycine dipeptide
molecule, and the C5 and C7
conformations have been identified as the ground and metastable conformations, 
respectively.
It has been confirmed that the substantial reduction of the free energy barriers,
thanks to the logarithmic form, 
enables us to efficiently reconstruct the free energy profile,
which was found to agree well with that 
obtained by the TI method. 
The free energy profile from the first-principles approach 
indicates that the empirical force field for the glycine dipeptide molecule
is sufficient to obtain the overall profile of the free energy landscape.

The LogMFD method allows us not only to easily sample rare events,  
but also to reconstruct the free energy profile ``{\it on-the-fly}" 
without suffering from the problems such as how to arrange the grid points 
or how to perform the numerical integration (as postprocessing) in TI.
It has been demonstrated in the present study that
free energy profiles using {\it ab initio} force field can be
reconstructed with less computational cost than is needed in
the TI method.
The FP-LogMFD method developed here is thus a promising tool for 
reconstructing free energy profiles, especially those
in which accurate descriptions for interatomic interactions are required.

\begin{acknowledgments}

The computation in this work was done using the facilities of 
the Supercomputer Center, Institute for Solid State Physics, 
University of Tokyo and the facilities of the Research Center for 
Computational Science, National Institutes of Natural Sciences, Okazaki, Japan. 
This work was partly supported by Grant-in-Aid for Scientific 
Research from JSPS/MEXT (Grant Nos. 22104012, 22340106, 23510120 and 24740297) 
and the Computational Materials Science Initiative (CMSI), Japan.

\end{acknowledgments}

\appendix
\section{\label{sec:appendix}Thermodynamic integration}

\begin{figure}
\includegraphics[width=7.5cm]{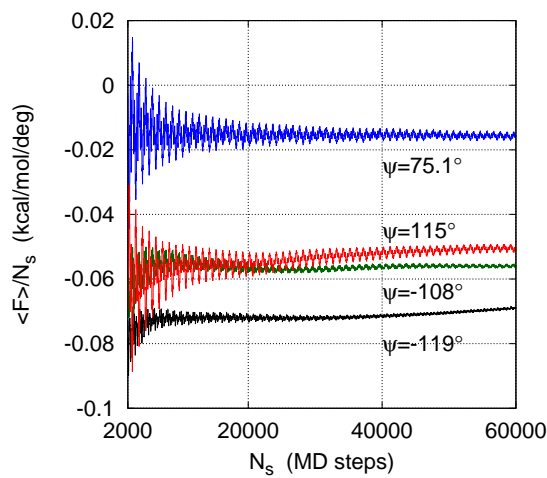}
\caption{\label{aveti} 
Cumulative averages of the mean force in the TI calculation
for several dihedral angles (
\textcolor{black}{
$\psi = -119^\circ, -108^\circ, 75.1^\circ, 115^\circ$})
under the condition 
\textcolor{black}{
$\phi=-80^\circ$}.}
\end{figure}

In order to check the result of the LogMFD calculations, 
thermodynamic integration (TI) was also performed 
for comparison
using the same computational conditions
for the first-principles MD calculations
(see Sec. \ref{subsec:parameter}). 
When one carries out a long-time CP-FPMD simulation, the energy tends to 
flow to the electronic degrees of freedom from the ionic degrees of freedom.  
Consequently, the lift from the BO surface of the electronic wave functions
becomes 
obvious, and finally, the simulation may break down.\cite{Pastore1991}
However, one can, instead, perform multiple short FPMD runs and
the mean force profile in the TI calculation can be constructed by 
averaging over the configurations from all of these short runs
at a given set of $\phi$ and $\psi$.
Figure~\ref{aveti} represents the convergence behavior of the MF
with several fixed $\psi$, as a function of the number of statistical samplings. 
\textcolor{black}{
From this result, 
}
we decided to use 60,000 FPMD steps in total
to estimate the MF
at each grid point of $\psi$ in our TI calculation.

\end{document}